Insights into Protein Unfolding under pH, Temperature, and Shear using Molecular Dynamics Simulations

Yinhao Jia[1], Clare Cocker,[2] Janani Sampath[1,*]

[1]Department of Chemical Engineering, University of Florida, Gainesville, FL, USA

[2]Department of Chemical Engineering, University of Virginia, Charlottesville, VA, USA

[*]jsampath@ufl.edu

**Abstract**

Protein biologics hold immense potential in therapeutic applications, but their ephemeral nature has hindered their widespread application. The effects of different stressors on protein folding have long been studied, but whether these stressors induce protein unfolding through different pathways remains unclear. In this work, we conduct all-atom molecular dynamics simulations to investigate the unfolding of bovine serum albumin (BSA) under three distinct external stressors: high temperature, acidic pH, and shear stress. Our findings reveal that each stressor induces unique unfolding patterns in BSA, indicating stressor-specific unfolding pathways. Detailed structural analysis showed that high temperature significantly disrupts the protein's secondary structure, while acidic pH causes notable alterations in the tertiary structure, leading to domain separation and an extended shape. Shear stress initially perturbs the tertiary structure, initiating structural rearrangements followed by a loss of secondary structure. These distinct unfolding behaviors suggest that different stabilization strategies are required to enhance protein stability under various denaturation conditions. Insights from these unfolding studies can inform the design of materials, especially polymers, aimed at improving protein stability.

**Introduction**

Proteins are used in many applications, ranging from food substitutes to therapeutics.[1–3] Their broad applicability makes their stability a central focus, as proteins are ephemeral and different stressors can easily perturb them during their lifetime, from synthesis to application. The resulting structural alternation brought about by these stressors leads to a decrease or loss of effectiveness in protein function.[4] These stressors include but are not limited to chemical environment change when a protein is synthesized and purified,[5] temperature fluctuations during storage and transport,[6] and shear stresses[7] (mechanical perturbation) while shaking, mixing, or injecting biologics. The diverse set of external stressors raises the question of whether these different stressors will cause the protein to unfold in different ways, as well as how distinct these unfolding pathways are.

Protein unfolding has been widely studied to determine intermediate states of the protein, which in turn could provide insights into the more complex folding phenomenon.[8] In addition, the unfolding, misfolding, and aggregation of proteins that manifest in several diseases, including Alzheimer's and Parkinson's, have been investigated to gain a deeper understanding of disease-causing mechanisms, which in turn help drug design.[9] In these studies, as well as many other experimental studies on protein unfolding, temperature is the most common external stimulus that has been used to induce and understand unfolding pathways of a wide range of proteins.[10] Techniques like differential scanning calorimetry (DSC) and circular dichroism (CD) give insights into the onset of unfolding, the corresponding structural change, as well as transition states. Recent research has demonstrated real-time tracking of protein unfolding with the combination of time-resolved X-ray solution scattering, allowing the observation of structural change in the protein as the unfolding progresses.[11] While a two-state model is usually applied to describe the folding/unfolding dynamics, a more recent Zimm-Bragg theory describes the unfolding process better.[12] The mechanism of chemical agent-induced denaturation has also been studied, however, observation of persistent residue structure even under high concentrations of chemical denaturants raises questions about the role of this stressor in destabilizing the protein.[13] Structural changes of protein under shear stress arising from fluid flow has long been debated, especially the role of the shear rate needed to unfold the protein. Some experimental studies have shown that there is no obvious structural change of proteins such as rhGH and IgG1 mAb at a shear rate of $10^4$ s$^{-1}$.[14] While high shear rates (~$10^7$ s$^{-1}$) were required to induce unfolding in proteins such as horse cytochrome c,[15] some studies report conformational perturbation even under low shear rates (~$10^2$ s$^{-1}$), in proteins such as lysozyme and bovine serum albumin.[16,17] These studies prove that proteins' structural response to stressors depends on the type of protein and stressor, and conclusions regarding the effect of one stressor cannot be applied to another stressor acting on the same protein.

While these experimental studies have uncovered the effect of a single stressor on protein unfolding, there have been few comparative studies that look into the differences between different stressors on a single protein.[18] Kishore et al. investigate the unfolding pathway of a multimeric chickpea β-galactosidase (CpGAL) under different conditions.[19] They found that chemical-induced denaturation causes complete unfolding and separation of the monomers, while pH-induced denaturation leads to structural relaxation and dissociation of monomers. In contrast, temperature-induced denaturation results in immediate aggregation following the initial unfolding. A more recent study by Kelly and Gage reported similar $\Delta G_{unfolding}$ values with slightly different rate constants while comparing thermal, chemical, and force unfolding in the muscle protein titin.[20] These studies provided a macroscopic view of the unfolding behavior. However, due to the temporal and spatial scales of experiments, probing microscopic details during the initial stages of protein unfolding, which are critical for understanding both the unfolding mechanism and protein folding, remains challenging. The complexity of capturing unfolding at short times confounds decoupling the influence of different stressors on protein structures using conventional experiments.

To capture protein unfolding at short times, molecular dynamics (MD) simulations not only offer atomic-level resolution of the unfolded protein, but also provide detailed insights into the mechanism of protein unfolding in response to different stressors. MD simulations have been used to capture the effect of single stressors on protein stability, giving insights into unfolding pathways.[21–25] Comparative studies between pH and thermal unfolding suggests diverged unfolding pathways for these stressors.[26,27] Gu et al. examined the human prion protein domain under different temperatures and pH, suggesting a secondary structure change accompanied by tertiary structure changes for pH, which was not observed under high temperatures.[26] Zhang et al. analyzed the unfolding behavior of a humanized antibody fragment (Fab) A33 under the impact of low pH and high temperature, observing different predicted aggregation-prone regions exposed to solvent and potentially lead to different aggregation mechanisms.[27] Languin-Cattoën et al. applied the Lattice Boltzmann Molecular Dynamics technique (LBMD) to investigate how a protein called CspA from E. coli., which consists of mainly β-sheet structure, unfolded under shear stress, and compared with force and thermal stress-induced unfolding.[28] Their findings emphasize that different stressors lead to distinct unfolding pathways. Notably, shear flow and thermal unfolding exhibited similar pathways, whereas force-induced unfolding was markedly different. While these studies provide valuable insights into how proteins respond to different stressors, they usually focus on small proteins and protein fragments. Larger proteins need to be simulated even though their size and complexity makes it difficult to do so. Proteins that are well characterized experimentally can be used to bridge the gap that might arise due to simulation size and timescales. Additionally, very few studies focus on the effect of flow-induced shear stress in protein stability, with even fewer comparative studies to distinguish the protein behavior under this stressor, and other stressors. Details regarding the unfolding pathway and weak regions of proteins that tend to fall apart first under different stressors is also lacking. Such information is critical not only for the problem of protein folding but also for customized design strategies to improve protein stability under specific external stressors.

Here, we conduct a systematic study on how three common external stressors namely, temperature, acidic pH, and shear stress denature bovine serum albumin (BSA), a well characterized protein which consists of α-helix bundles and a defined tertiary structure. We provide a comprehensive picture of the microscopic response of BSA to the different stressors and its unfolding pathway. Global structure metrics suggest different behavior of BSA when subject to different external stimuli. Analysis on the domain level indicates divergent BSA unfolding pathways as a result of different stressors. The free energy landscape of proteins is constructed using the first two principal components and shows a broadening of the free energy minima under external stressors. The different unfolding pathways suggest that the different stressors disrupt the protein structure in unique ways.

**Methods**

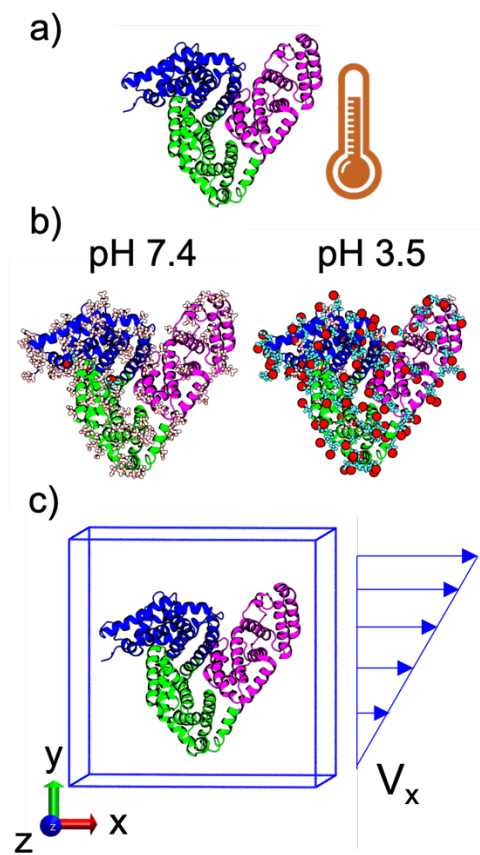

**Figure 1: Schematic of system and setup under different external stressors, with (a) high temperature, (b) acidic pH, and (c) shear stress. The three colors represent the three domains of BSA, where blue, green and magenta indicate Domains I, II and III, respectively.**

**System Setup and Unfolding Protocol**

The structure of bovine serum albumin (BSA) was obtained from protein databank (code: 4f5s). The protonation states of titratable residues were determined using the PDB2PQR webserver[29] at both pH = 7.4 and 3.5. There are 116 titratable residues in BSA, namely 40 ASP, 59 GLU, and 17 HIS residues. At pH 7.4, no ASP or GLU residues are protonated, and only 3 out of

17 HIS residues are protonated (both *pros*- and *tele*-nitrogen are protonated). However, at pH 3.5, 30 out of 40 ASP residues are protonated, with HD2 linked to OD2 at the end of the side chain, contributing to a net +30 charge. Additionally, 55 out of 59 GLU residues are protonated, with HE2 linked to OE2 at the end of the side chain, contributing to a net +55 charge. An extra 13 HIS residues are protonated, with H241 remaining half-protonated on *pros*-nitrogen, contributing to a net +13 charge. The change in residue protonation state results in 98 additional charges, shifting the net charge of the protein from -14 to +84. These protonation states slightly differ from a previous study, which reported a total charge of -9 at pH 7.4 (with 8 HIS residues protonated) and a total charge of +100 at pH 3.5 (with all ASP, GLU, and HIS residues protonated).[22]

Simulations with varying pH and temperature conditions were conducted using GROMACS 2023.2 software[30] with GPU acceleration. CHARMM36m[31] force field was used to model the protein together with TIP3P water. The protein was solvated in a cubic box with the edge at least 2.5 nm from the protein, resulting in cubic box sizes ranging from 13.9 to 14.2 nm with a total of ~ 260,000 to ~ 290,000 atoms. $Na^+$ or $Cl^-$ ions were introduced into the simulation box to neutralize each system. Hydrogen bonds were constrained during the simulation using the LINCS[32] algorithm. The cutoff for short-range electrostatic and Van der Waals interactions was set to 1.2 nm. The particle-mesh Ewald[33] method was used to compute long-range electrostatic interactions; fast Fourier transform grid spacing was set to 0.15 nm. Van der Waals interactions were smoothed between 1.0 to 1.2 nm using a force-base switching function. Constant temperature was maintained using the Nosé – Hoover[34] algorithm. For the constant-pressure runs, the Parrinello-Rahman[35] barostat (1 bar, 2.0-ps coupling constant) was used. The time step for numerical integration was 2 fs.

The systems, prepared as described above, were first subject to steepest descent energy minimization with a maximum force of 1000 kJ·mol$^{-1}$·nm$^{-1}$, followed by a stage of 100 ps NVT equilibration. In this stage, heavy atoms of the protein were restrained, and the solvent molecules were equilibrated. Subsequently, a 100 ps constant pressure and temperature (NPT) simulation was conducted at 310 K to equilibrate the system with position restraint on the protein heavy atoms in place. The position restrain was then lifted, and the outputs were used as a starting point for the production simulation. For the assessment of acidic pH induced unfolding, the protein structure prepared at pH 3.5 was simulated at 310 K. For the assessment of protein unfolding at high temperature, the system prepared at pH 7.4 was simulated at both 310 K and 500 K, following a previously established protocol.[25] Three replicates of each system were performed to increase reproducibility and reliability. The starting configuration of each replica was randomly selected from the trajectory of the first simulation performed at 310 K and pH = 7.4.

LAMMPS[36] was used to assess protein unfolding under shear stress. Couette flow is applied to the simulation box along the xy plane (Figure 1c), which can be envisioned as fixing the bottom of the cubic simulation box while moving the top towards the x direction, which will generate a linear velocity gradient from the top to bottom. This was implemented using the *fix deform* command with the option *remap v*. To account for changes in atom positions with changing box shape the *fix nvt/sllod* command was used. The magnitude of the induced shear stress was controlled by tuning the *erate* or engineering shear strain rate. The shear stress ($\tau$) can be calculated according to equation (1), with the velocity of water (u) at the top of the simulation box and the plate separation (h) determining the shear rate ($\dot{\gamma}$). To optimize the magnitude of shear stress needed to induce protein unfolding within a reasonable simulation time, we tested shear rates ranging from 1×10$^{-5}$ to 5×10$^{-5}$ fs$^{-1}$ (Figure S1). We chose a shear rate of 3×10$^{-5}$ fs$^{-1}$ to assess the unfolding behavior of BSA, as this gave an optimal balance between computational resources and unfolding time. Two repulsive walls were placed at the lower and upper bounds of the y-axis of the simulation box. This ensures that the protein molecules do not cross the boundary when experiencing an instantaneous velocity change.

$$\tau = \eta\dot{\gamma} = \eta\frac{\partial v_x}{\partial y} = \eta\frac{u}{h} \quad (1)$$

The starting conformation for the shear run was taken from the NPT equilibrated system at 1.0 bar, 310 K and pH = 7.4. The GROMACS format topology was processed using ParmED[37] and converted into LAMMPS topology using the *ch2lmp* tool available in LAMMPS. The force-switching function was used to represent the van der Waals interactions. The cutoff for short-range electrostatic and van der Waals interactions was set with an inner and outer cutoff of 8 Å and 12 Å, respectively. Nosé – Hoover thermostat was used to maintain a constant temperature of 310 K with a damping factor of 200 fs. The time step for numerical integration was 2 fs. The initial velocity was generated according to Gaussian distribution using random seed. To accelerate the formation of an equilibrated velocity profile that matches the rate of box deformation, the initial velocity was mapped based on the location of a particular atom utilizing the *velocity ramp* command. To monitor the velocity profile across the simulation, the box was divided along the y-axis into 20 layers, and for each layer, the velocity of water along the x-axis was calculated by averaging the velocity of the atoms every 2 ps, over a period of 0.2 ps to improve statistics. The velocity profile can be found in Supporting Information. (Figure S2)

The protein was rotated 90° around the y-axis or z-axis to generate two other distinct configurations (Figure S3), so that different parts of the protein are subjected to shear, resulting in three different configurations in total, allowing the examination of how different directions of shear stress will induce unfolding. Additionally, each configuration was run for 3 replicates with different initial velocity seeds to improve reproducibility.

**Analysis**

Simulation trajectories were analyzed using GROMACS tools and visualized using VMD. The predicted circular dichroism (CD) spectra were generated by the PDBMD2CD[38] Colab version using the simulation trajectories. Protein structures under various stressors were analyzed at the point when their RMSD first reached 1.0 nm, while the structure of the baseline system at 310 K and pH 7.4 was taken from the last frame of each replicate. The secondary structure of the protein was assigned using the inbuilt timeline tool of VMD, with the STRIDE algorithm. Root mean square deviation (RMSD) of the whole protein was calculated by fitting to the backbone of the initial structure, whereas the per domain RMSD was calculated by fitting to the domains' initial backbone structure during the production run.

Residue contact analysis was carried out using the Python package Contact Map Explorer. Contacts between residues were determined based on distances, with a cutoff of 4.5 Å for heavy atoms. For the protein at physiological temperature and pH, contacts were calculated from the final 10 ns of each simulation trajectory. Contacts within unfolded proteins at 500 K pH 7.4 and 310 K pH 3.5 were calculated over a 10 ns window after the RMSD of BSA first reached 1.0 nm. For the systems subjected to shear stress, residue contacts were calculated within a 0.35 ns window before and after the RMSD reached 1.0 nm. The results were averaged over three independent replicates.

Principal component analysis (PCA) was conducted to construct free energy surface of protein unfolding under different external stressors. For each stressor, replicate simulations are first concatenated into a single long trajectory. Then, a covariance matrix of the Cα atoms was constructed by fitting the protein trajectory to the reference structure of protein backbone atoms using least-squares fitting. The resulting covariance matrix was diagonalized to obtain the eigenvectors as well as the corresponding eigenvalues using *gmx covar*. Each eigenvector represents a principal component (PC) that depicts a mode of movement, with the corresponding eigenvalue indicating the extent of the motion along the eigenvector. The first two PCs, PC1 and PC2, have the highest dynamic variations, and are used to construct the free energy surface, where the simulation trajectories were projected to the 2D space of PC1 and PC2 using *gmx anaeig*. The projection was processed to construct the free energy surface using *gmx sham*.

**Results**

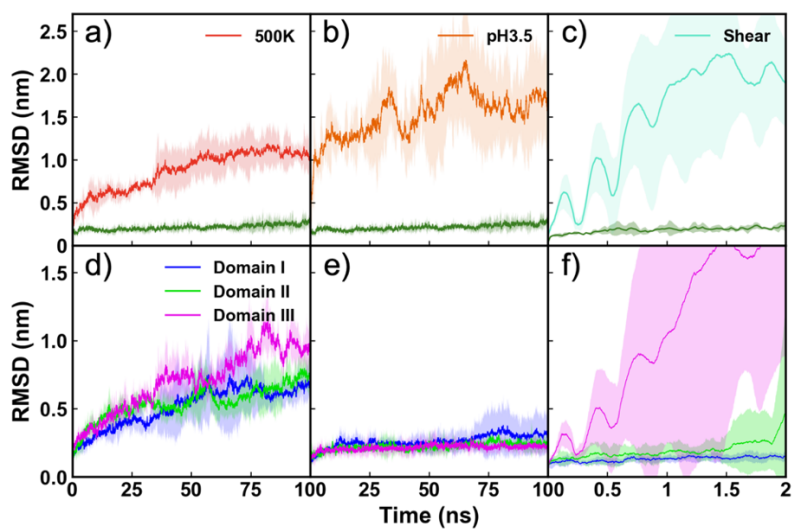

**Figure 2: Backbone root mean square deviation (RMSD) of the whole BSA protein (a, b, c) and its domains (d, e, f) at (a, d) 500 K (b, e) pH=3.5 and (c, f) shear stress. The dark green line represents baseline results at 310 K pH 7.4 and 0 shear stress. Domain colors are consistent with the protein in Figure 1. Solid lines indicate the mean values averaged over three independent trials, while the shaded regions show one standard deviation. We note that**

**the maximum time for the shear plots is 2 ns, whereas that for the high temperature and pH plots are 100 ns.**

To capture the change in BSA structure, we calculate the root mean square deviation (RMSD) as described in the Methods section. We find that BSA at 310 K, pH = 7.4, and 0 shear is stable, as seen in Figures 2a, 2b, and 2c where the RMSD is a steady horizontal line across the simulation duration, indicating little to no alteration of the BSA structure. This is also consistent across the simulations performed on GROMACS and LAMMPS. We compare change in BSA structure under different stressors to the baseline system. As shown in Figure 2a, high temperature induces a considerable structural change in BSA, indicated by the steep increase of RMSD beyond 5 Å in less than 5 ns, which steadily increases for the rest of the simulation time, resulting in the RMSD going beyond 1 nm at the end of the 100 ns simulation. The RMSD change of BSA at pH 3.5 paints a different picture (Figure 2b). At the start of the simulation, the RMSD undergoes a dramatic rise within 2 ns and attains a value greater than 1 nm, after which the RMSD steadily increases, reaching values near 2.0 nm at the end of the simulation. This suggests that the protein experiences greater structural change due to acidic pH than high temperature. When BSA is subjected to shear stress, the RMSD exhibits a sinusoidal response with an overall increase with time (Figure 2c). This indicates that while the structure of BSA was perturbed under Couette flow, it recovers some of the structure partially at very short times. However, as the flow progresses and the flow field develops, it eventually disrupts the entire structure beyond recovery. A similar pattern was observed in other studies that conducted protein simulations under shear stress, where the radius of gyration ($R_g$) was monitored, and periodic fluctuations along with an overall increase in $R_g$ were observed.[28] Due to the relatively high shear stress applied here, the interval for data collection is smaller so that the changes can be clearly observed. The trend in RMSD indicates different unfolding mechanisms of BSA under shear stress, compared to temperature and pH.

To better understand structural changes and to identify how the three stressors perturb BSA at a more local level, we conducted RMSD analysis on the three domains, shown in Figure 1. This was calculated by performing a root mean square fitting on the backbone atoms of a specific domain to its initial structure. The domains (Domain I: residues 1 to 196, Domain II: residues 204 to 381, and Domain III: residues 382 to 583) are assigned according to the BSA sequence deposited in the CATH database. As expected, no apparent structural alteration was observed in the baseline system (Figure S1). At 500 K (Figure 2d), all domains exhibit a consistently increasing trend in RMSD, with Domain III reaching the highest RMSD of 1.0 nm, indicating that all domains undergo significant structural changes at high temperatures. Under acidic pH conditions (Figure 2e), the RMSD change of each individual domain is ~2.5 Å, close to the RMSD at pH 7.4, which fluctuates around 1~2 Å over 100 ns (Figure S4). This implies that while acidic pH results in significant structural alternation of the entire protein, the individual domains themselves do no change significantly. While this might seem contradictory, we predict that the domains are moving apart from one another at low pH (high inter-domain motion), while retaining stability at the intra-domain level. Under shear stress (Figure 3f), we find that Domain III exhibits the most drastic change, similar to temperature. The RMSD for Domain III under shear stress shows a similar pattern to that observed at the whole protein level: an initial sinusoidal fluctuation followed by steady elevation. The structural changes in Domains I and II are considerably lower than those in Domain III, however, Domain II shows an increase later in the simulation.

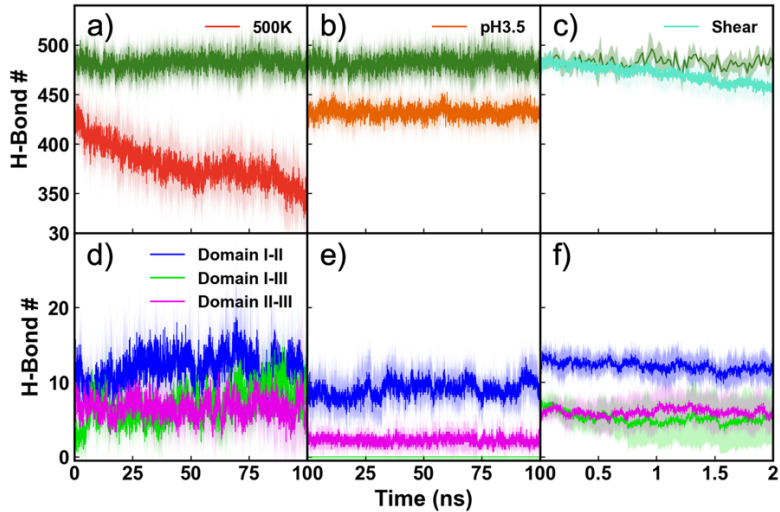

**Figure 3: Number of hydrogen bonds in the whole BSA protein (a, b, c) and between pairs of domains (d, e, f) under (a, d) high temperature, (b, e) acidic pH, and (c, f) shear stress. The dark green line represents baseline results at 310 K pH 7.4 and 0 shear stress. Domain colors are consistent with the protein in Figure 1. The solid lines indicate the mean values averaged over three independent trials, while the shaded regions show one standard deviation. We note that the maximum time for shear plots is 2 ns, whereas that for the high temperature and pH plots are 100 ns.**

To understand the interactions that breakdown as BSA unfolds, we examine the evolution of the number of hydrogen bonds under the influence of the three external stressors, as shown in Figure 3. The baseline protein displays a steady hydrogen bond network over the simulation time, with an average total of ~ 480 (±10) hydrogen bonds. At 500 K (Figure 3a), BSA gradually loses hydrogen bonds, resulting in ~350 (±19) hydrogen bonds at the end of the 500 K simulation. For the acidic pH system, we find that the number of hydrogen bonds is lower than the baseline to begin with; ~430 (±10), and it fluctuates around this value over the duration of the simulation. The change of the protonation state of the titratable residues likely induced a loss of hydrogen bonds from the start, resulting in a difference of about 50 hydrogen bonds compared to the baseline protein. We find that shear stress also leads to a loss of hydrogen bonds (Figure 3c) in a similar fashion to temperature. However, the hydrogen bonds lost is not significant compared to the other two conditions due to the short timescales considered. When extending the analysis to a total time of 6 ns, we find that the hydrogen bonds have reduced to ~415 (±55).

In addition to the total number of hydrogen bonds in the protein, intra-domain and inter-domain hydrogen bonds between pairs of domains are also investigated (Figures 3d-f, S5). With high temperatures, the loss of intra-domain hydrogen bonds was observed in all three domains (Figure S5). In contrast, the inter-domain hydrogen bonds do not show a drop (Figure 3d). Specifically, the number of hydrogen bonds formed between Domains I and II and Domains II and III do not change much compared to the baseline protein, and more hydrogen bonds are observed between Domains I and III. Such a trend suggests Domain I and Domain III come close to each other at high temperatures. With the acidic pH, the number of intra-domain hydrogen bonds follows a similar trend as that of the total number of hydrogen bonds, where after the initial drop, the number of hydrogen bonds fluctuates without significant change for the rest of the simulation (Figure S5). As for the inter-domain hydrogen bonds (Figure 3e), Domains I and II and Domains II and III are very similar to the baseline, and do not change much over the simulation duration. However, the hydrogen bonds between Domains I and III are almost entirely lost, suggesting a large separation leading to a loss of contact between these two domains. Shear stress also induces the loss of intra-domain hydrogen bonds (Figure S5), but those belonging to Domain I are largely not influenced. As was the case with acidic pH, the inter-domain hydrogen bonds between Domains I and II and Domains II and III were not significantly influenced, however, hydrogen between Domains I and III were reduced.

The change of hydrogen bonds formed in the whole protein, intra- and inter- domains all suggest distinct behavior of proteins under different stressors. These results also indicate which parts of the protein are more susceptible to a particular stressor. At high temperatures, the overall loss of hydrogen bonds with no significant change in inter-domain hydrogen bonds indicates that the secondary structure of the protein is likely collapsing. In contrast, loss of inter domain hydrogen bond under acidic pH conditions suggests the possibility of domain separation between Domains I and III. Of the three domains, Domain III appears to be most vulnerable to shear stress.

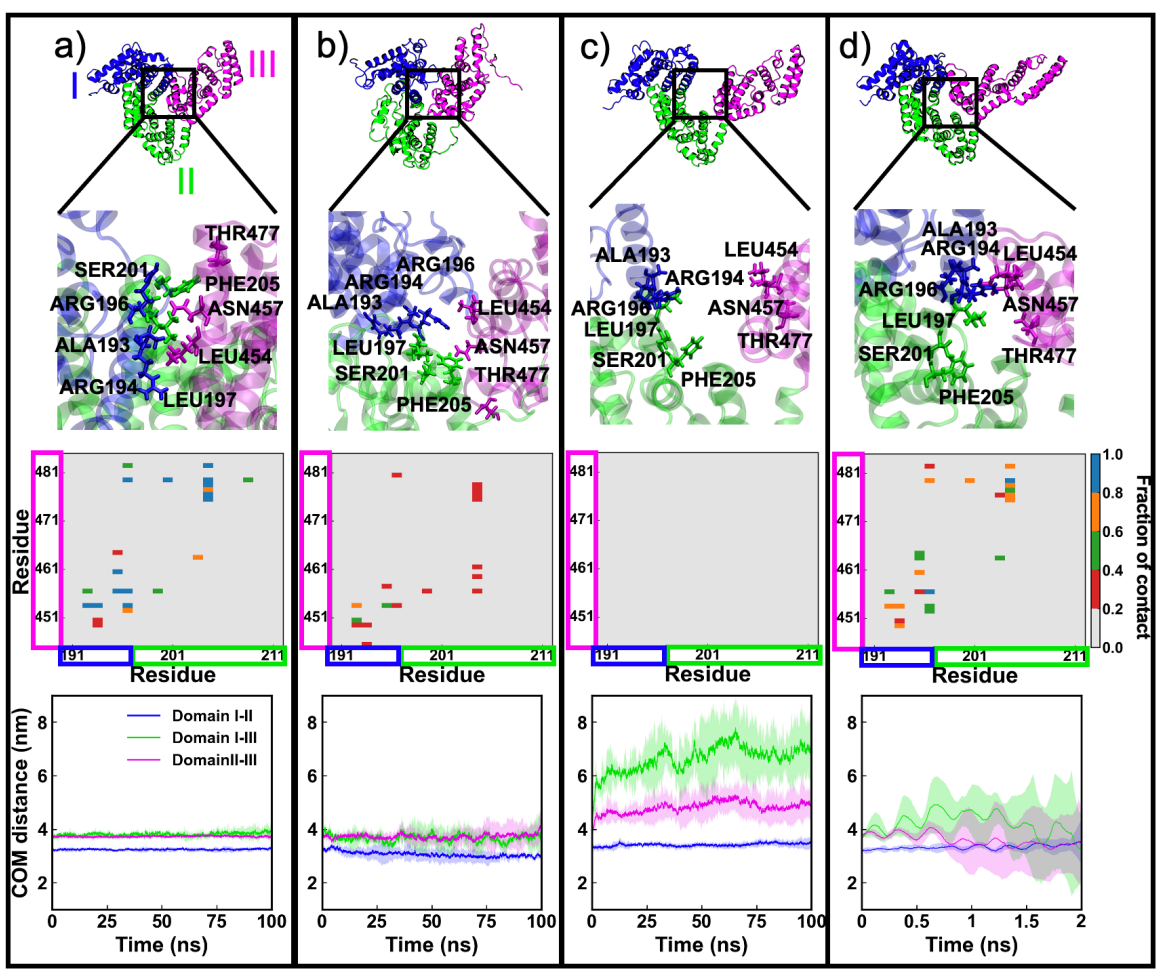

**Figure 4:** Conformation analysis of BSA under different stressors, with (a) 310 K pH=7.4 and no shear stress, (b) 500 K pH=7.4, (c) 310 K pH=3.5, and (d) 310 K pH=7.4 shear rate 5×10⁻

**5 fs⁻¹. Each panel from top to bottom includes a snapshot of BSA at RMSD = 1.0, a snapshot of residue contact, residue contact analysis, and center of mass distance between different domains. The boxes around the reside axes in the contact map analysis indicates the domain that each residue belong to.**

To better understand the results from the RMSD and hydrogen bond analyses, we capture snapshots and calculate residue contact maps of BSA under different stressors and compare them to the baseline system. These results are shown in Figure 4. The snapshots are representative and are captured when the total protein RMSD (Figure 2) reaches 1 nm for each stressor. Details about contact map calculation can be found in Methods section. We find that each stressor causes a different conformational change in the protein, even though the overall extent of structural deviation is the same in all systems (same RMSD value). For the baseline conditions, BSA presents a 'heart' shaped structure (Figure 4a), and the contact analysis indicates persistent contact (contact fraction > 0.8) between Domains I and III and Domains II and III (Figure 4a), where the residues ARG 196 and SER 201 are located on the same alpha-helix structure connecting Domain I and Domain II. Although these contacts are less persistent at high temperatures (Figure 4b), the inter-domain contacts still remain. From the snapshot (Figure 4b), we see that high temperature results in helices losing their secondary structure and becoming flexible loops, while the domains are still close to each other. This is consistent with the inter-domain hydrogen bond result (Figure 3d). For BSA subjected to acidic pH, a clear separation between the domains is observed in the snapshot (Figure 4c), and the original 'heart' shape appears extended, which was reported previously to be the conformational transition from N-form to F-form.[39] The contact map shows a complete loss of interdomain contacts in acidic BSA (Figure 4c). This is because the acidic pH changes the protonation state of a large number of residues, which results in significant positive charge on the surface; electrostatic repulsion drives the domain separation and the extension of BSA.[22] However, unlike temperature, we find that secondary structure of the protein is preserved under acidic pH conditions. We find that shear stress also induces elongation of the BSA structure (Figure 4d), and this is most pronounced in Domain III. We also find that the contact between Domains I and III is reduced but not entirely lost (Figure 4d). Applied shear stress pulls a part of the helix bundle in Domain III which results in the loss of tertiary structure.

We track the evolution of distance between different domains during the unfolding process by examining changes in the center of mass distance between the domains. At physiological temperature and pH (Figure 4a), the center of mass (COM) distances between different domains remain consistent with minor fluctuation. With high temperatures (Figure 4b), the COM distance between Domain I and II is only slightly lower than that at 310 K. The distance between the centers of mass of Domain I and III and Domain II and III do not change appreciably. Under acidic pH however, we find a large difference interdomain motion. The distance between Domains I and II is the lowest, suggesting that they do not move apart due to lowering of pH. However, Domains I and III and Domain II and III undergo a significant increase when compared to the baseline condition, suggesting that Domain III is likely the cause of instability at low pH. Previous simulation studies also reported the separation of Domain I and III.[22] For shear stress, the COM distance between Domains I and II was not very large during the simulation periods. However, the distance between Domains I and III, and Domains II and III show large amplitude fluctuations, indicating that the domains are likely moving apart and coming closer together due to tumbling motion of the protein as it is being sheared. Such change in the domain-wise distance suggests that the pH-induced electrostatic change to the protein surface leads to the separation of domains, distinct from how temperature unfolds BSA without affecting the inter-domain, as well as how shear induces domain structure rearrangement in BSA.

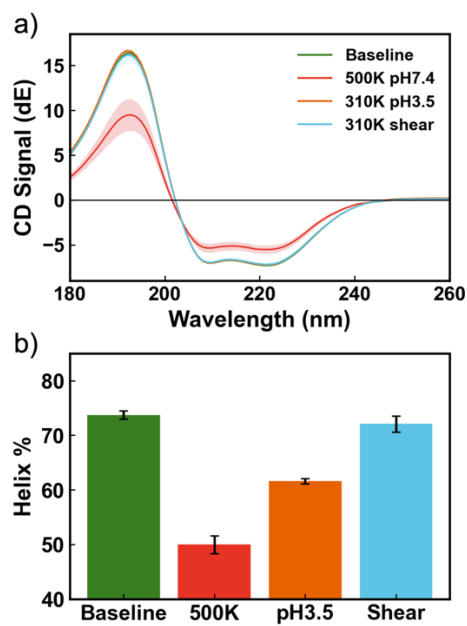

**Figure 5: Secondary structure analysis using (a) simulated Circular Dichroism and (b) VMD under different stressors, with structure taken from when the protein RMSD first reaches to 1.0 nm.**

To quantify the structural change in BSA and to compare with prior experiments, we calculate the circular dichroism (CD) spectra from the simulation trajectories and quantify the percentage of α-helices after BSA is subjected to external stressors. To make a consistent comparison, we perform these calculations when the protein RMSD first reaches 1.0 nm. The baseline structure is obtained from the last 2 ns of each replicate. From Figure 5a, the simulated CD spectrum reveals that only high temperature significantly alters BSA's secondary structure, with a lowering of the first peak at 190 nm, as well as a minima observed at longer wavelengths.

The positive band at 193nm and negative band at 222 nm and 208 nm present the α-helical structure of the protein.[40] The lowering of the peak at 193, 208, and 222 nm indicates the loss of α-helical in BSA. In contrast, pH and shear do not change the CD signature as much, presumably because CD cannot capture changes to the tertiary structure of the protein. Experimental studies reported secondary structure loss of BSA at high temperature and under shear flow.[17,41,42] Additionally, previous simulation study on low pH induced BSA unfolding also reported the loss of its α-helical structure.[22] Results from the secondary structure analysis is shown in Figure 5b. At 310 K and pH 7.4, 74% of BSA residues form a helical structure; this is slightly higher than experimentally observations of 67% at 20°C.[39] High temperature induces structural changes, reducing the helical structures to 50%. At acidic pH of 3.5, helical structure accounts for 62% of BSA, lower than that at physiological pH. Under shear stress, 73% of BSA maintained a helical structure, showing minimal change compared to the native state. This confirms that the large change in RMSD is not due to loss of secondary structure but due to changes in the proteins' tertiary structure.

    The secondary structure analysis in conjunction with the domain motion suggests how BSA unfolds under these three stressors. With high temperature, the protein secondary structure is lost, resulting in the denaturation of the protein. Under acidic pH, while there is a loss of secondary structure evidenced by the alpha-helix fraction, the most significant change is to tertiary structure of the protein, as seen in the domain COM plot. Shear stress perturbs the secondary structure of the protein the least. It unfolds the protein by causing the domains to rotate and tumble, disrupting the tertiary structure. This is consistent with the observation in previous experimental studies where the change in tertiary structure occurs in low shear range, and the disruption of the three-dimensional conformation of native BSA precedes the unfolding of helices.[17] We might observe

changes to the secondary structure if we extended the size and time of our simulations, however, we are interested in the initial unfolding of the protein.

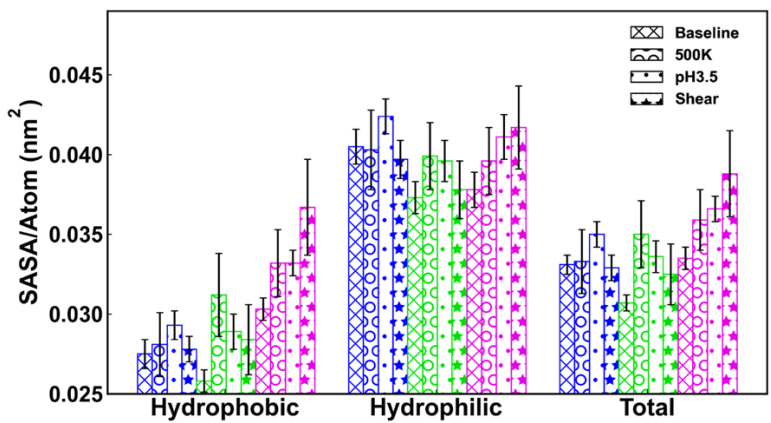

**Figure 6: Solvent accessible surface area (SASA) per atom of BSA domains under different external stressors, with histograms show mean values averaged over three independent trials and error bars represent one standard deviation. Blue, green, and magenta plots represent Domains I, II, and III, respectively.**

The solvent-accessible surface area (SASA) was calculated for the entire protein (Figure S6) as well as each domain (Figure 6) to understand the change in protein surface to different stressors, as a function of hydrophobicity. At the protein level, SASA first fluctuates at the same baseline value, and then steadily increases. At acidic pH, the increase in SASA happens instantaneously, after which it fluctuates around the same level for the remainder of the simulation. Shear stress induces a steady increase of SASA from the start of the simulation, and shows a sinusoidal fluctuation similar to that observed in RMSD and HBN. The SASA/atom was calculated after the BSA structure first reaches 1.0 nm and averaged over 10 ns. Due to the relatively limited time frames available in the shear simulation, we calculated SASA/atom for 0.35 ns before and after the RMSD reaches 1.0 nm. As shown in Figure 6, the extent of change of each domain to the stressors is different. For Domain I, acidic pH results in a higher increase of SASA than other conditions, regardless of hydrophobicity. Domain II experiences higher SASA under high temperature, followed by acidic pH and then shear, regardless of hydrophobicity. Domain III, however, experienced the highest SASA/atom increase from the shear stress, followed by acidic pH and high temperature. Such a different trend suggests that different domains of the protein have different sensitivities to the external stressors' influences, even though they all contain alpha helix bundles. Overall, high temperature influences Domain II, acidic pH influences Domain I, and shear stress influences Domain III. We note that the domain SASA result obtained for acidic pH is slightly different from previous simulation study, which could be due to the different ways in which hydrophobicity is defined, as well as different protonation states.[22]

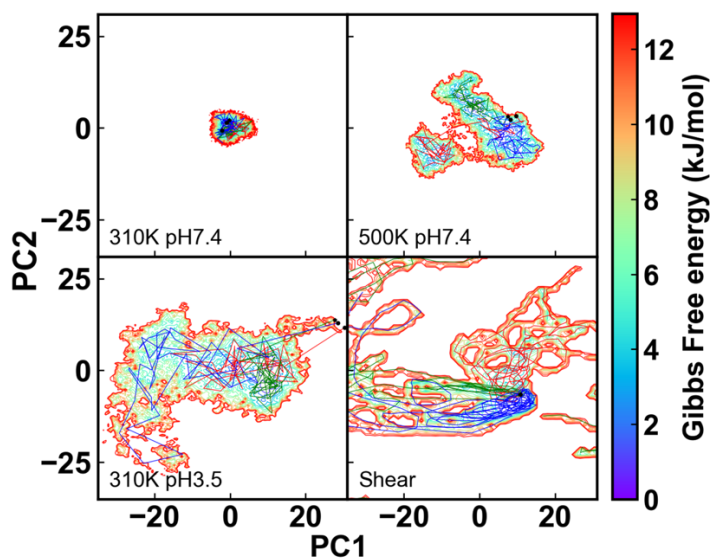

**Figure 7: Free energy surface constructed using principal components obtained from PCA, with line plots in different colors indicating the progress of each simulation replicates moving through the free energy surface.**

Finally, to capture the thermodynamics of the unfolding process, we performed principal component analysis (PCA) and use the first two principal components to construct the free energy surface for each external stressor. As seen in Figure 7, at the physiological temperature and pH, the protein is restricted to a steep and narrow free energy basin, signifying that BSA cannot unfold readily under these conditions. At high temperatures, BSA is pushed out of its native state and samples a broader range of configurations which includes many metastable intermediate unfolded states. At low pH, due to inherent limitations of classical MD simulations, the protein was in a high energy state at the beginning of the simulation as the protonation state of the residues was fixed before performing simulations. We find that BSA spontaneously moves into a more favorable the low energy state, where the free energy surface shows a variety of intermediate unfolded states of BSA. We also notice that the transition of BSA from the initial high-energy state to the low-energy state follows the direction of the positive PC1 and PC2 towards the negative part. Shear stress, however, is similar to temperature, where the protein is forced to move out of the native energy well due to force from shearing the system, and visits many other configurations, thus broadening the free energy surface. We find that excessive energy conferred on the system from external stressors forces BSA out of its well-folded native state to visit other states through distinct pathways, unique to each stressor.

**Conclusion and Outlook**

By conducting MD simulations under various conditions - high temperatures, acidic pH, and shear stress, we are able to capture how different external stressors unfold BSA. Structural analysis at the entire protein level, including backbone RMSD, the number of intra-protein hydrogen bonds, and solvent-accessible surface area (SASA), indicate different onset times of unfolding, suggesting different stressors unfold the protein differently. Further analysis at the domain level, both inter- and intra- domain, reveals specific mechanisms of unfolding under various stressors. High temperatures break hydrogen bonds within each domain, increasing the surface area exposed to water and causing a large structural change in each domain, leading to the loss of secondary structure. However, the distance between domains changes only slightly. Under acidic pH, significant changes in electrostatic interactions occur, converting BSA from slightly negatively charged to significantly positively charged. These electrostatic changes induce minor structural changes at the domain level, with some loss of secondary structure, but lead to dramatic changes between domains. There is a net loss of contact between Domains I and III, as well as between Domains II and III, leading to domain separation and change of tertiary structure. Shear stress affects Domain III the most, where significant structural perturbations are observed, although the loss of secondary structure is not apparent even in Domain III. The dominant effect of shear stress on the protein structure is the spatial rearrangement of the helical structure, resulting in changes to BSA's tertiary structure, which causes an eventual loss of secondary structure over a longer period.

Given the delicate nature of proteins and the various stressors they encounter throughout their lifespan — from synthesis to application — it is crucial to incorporate knowledge of how proteins destabilize when engineering them for non-native conditions. For example, high temperatures can cause a protein's secondary structure to unfold, whereas pH changes can alter its tertiary structure, as observed in BSA. By strategically designing mutations or attaching polymers such as polyethylene glycol (PEG), depending on the environment it will be used in, materials can be engineered to stabilize protein structures. Understanding where and how unfolding occurs can inform this design process. Previous research has demonstrated that using surfactants like SDS can enhance BSA stability at high temperatures by facilitating interactions between SDS ions and BSA's helical regions.[43] Additionally, conjugating pH-responsive polymers like pDMAEMA has been reported to enhance protein stability in acidic conditions.[44] Polymer conjugation can also shield the protein from mechanical stress. Polyethylene glycol, for example, has been shown to extend the lifetime of alpha-helices under constant stress.[45] Future investigations will explore how proteins respond to combined stressors, such as thermo-mechanical stress. A clear understanding of how a protein denatures can guide research and optimization of conjugation sites and materials to enhance protein stability in a rational manner.

**Acknowledgement**

J.S acknowledges startup funds provided by the Department of Chemical Engineering and the Herbert Wertheim College of Engineering at the University of Florida (https://www.che.ufl.edu). J.S and Y.J gratefully acknowledge funding provided by the Oak Ridge Associated Universities (ORAU) Ralph E. Powe Junior Faculty Enhancement Award. C.C acknowledges support by the National Science Foundation under Grant No. 1852111. Y.J and C.C acknowledge University of Florida Research Computing for providing computational resources and support (https://www.rc.ufl.edu/).

**Supporting Information**

BSA structure analysis under different shear rate, velocity profile of water during shear stress induced unfolding simulations, starting conformation of BSA experience shear stress from different directions, RMSD of BSA domains under physiological conditions, number of intra-domain hydrogen bond of BSA and total SASA of BSA under different conditions, and cumulative contribution of all principal components.

Supporting information for

Insights into Protein Unfolding under pH, Temperature, and Shear using Molecular Dynamics Simulations

Yinhao Jia[1], Clare Cocker,[2] Janani Sampath[1,*]

[1]Department of Chemical Engineering, University of Florida, Gainesville, FL, USA

[2]Department of Chemical Engineering, University of Virginia, Charlottesville, VA, USA

[*]jsampath@ufl.edu

Table of contents

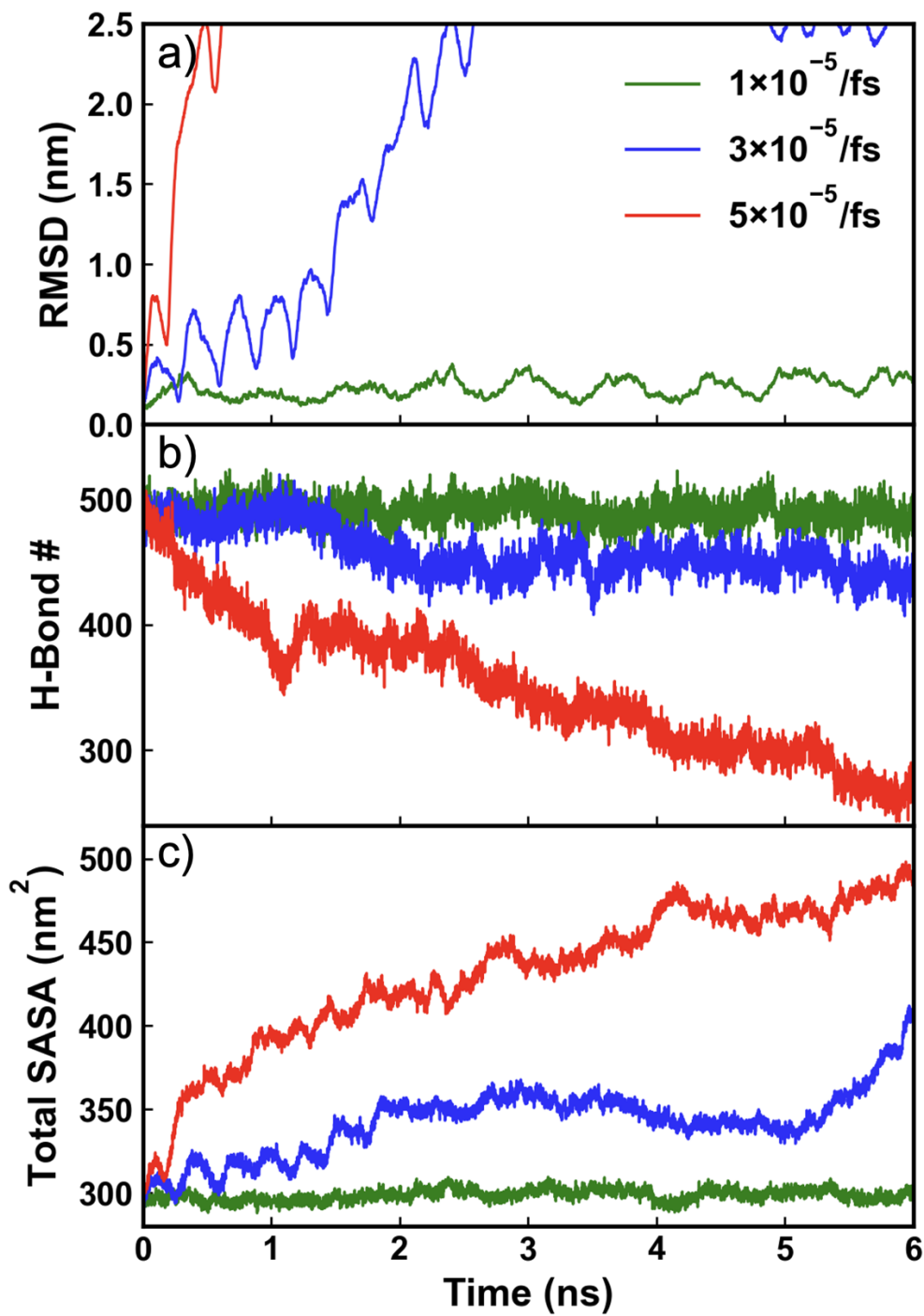

**Figure S1:** BSA structure analysis of (a) RMSD (b) total number of hydrogen bond (c) Total solvent accessible surface area (SASA) under different shear rate.

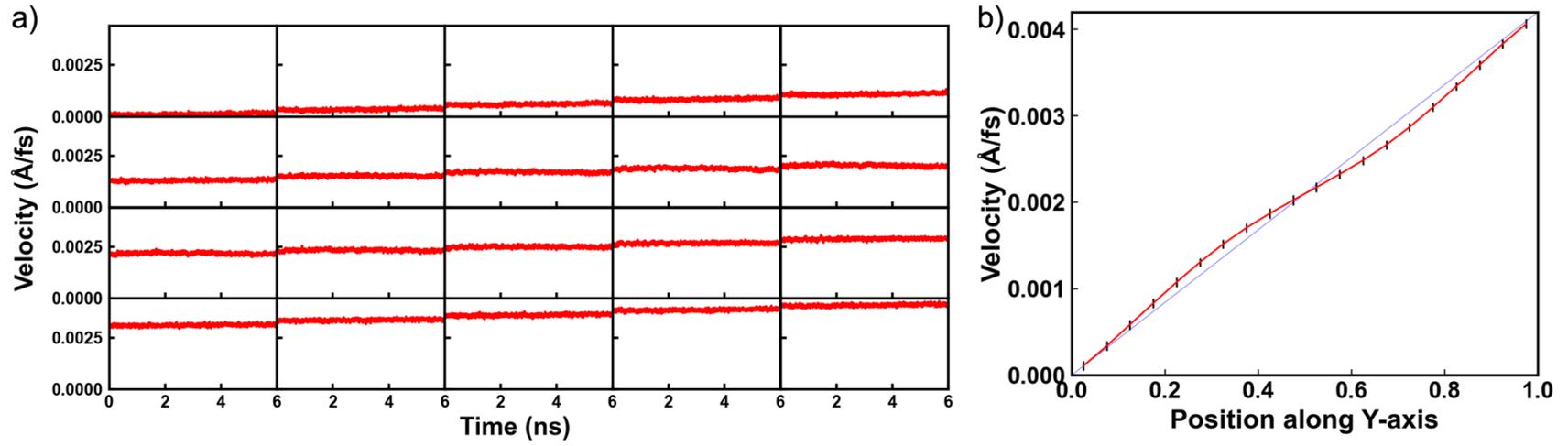

**Figure S2:** (a) Velocity profile of each layer along the y-axis and (b) average velocity profile during the simulation

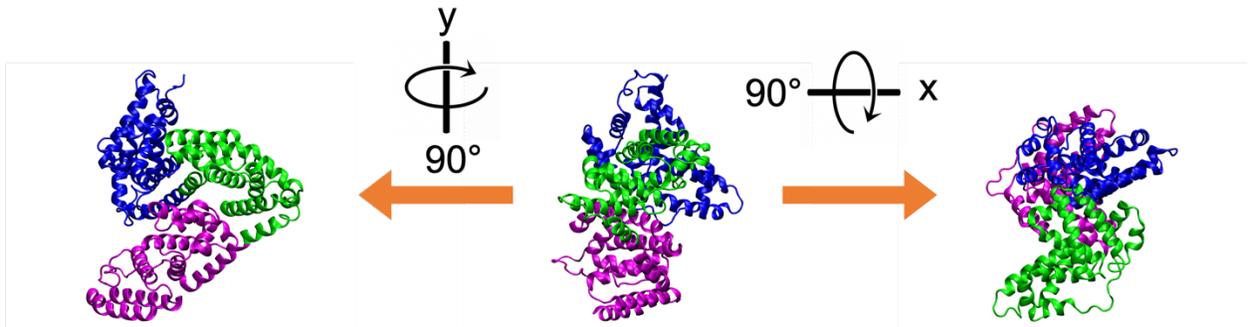

**Figure S3: conformation of BSA experience shear stress from different directions**

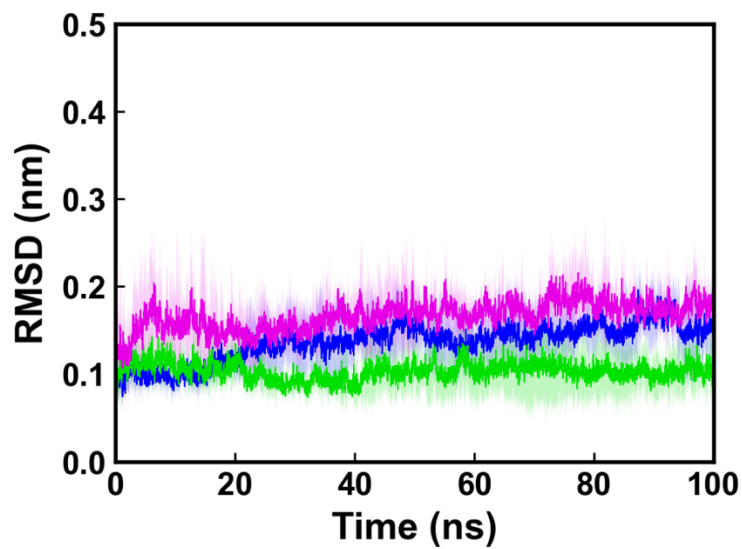

**Figure S4: Backbone root mean square deviation (RMSD) of the BSA domains at 310 K. Domain colors are consistent with the protein in Figure 1. The solid lines indicate the mean values averaged over three independent trials, while the shaded regions show one standard deviation.**

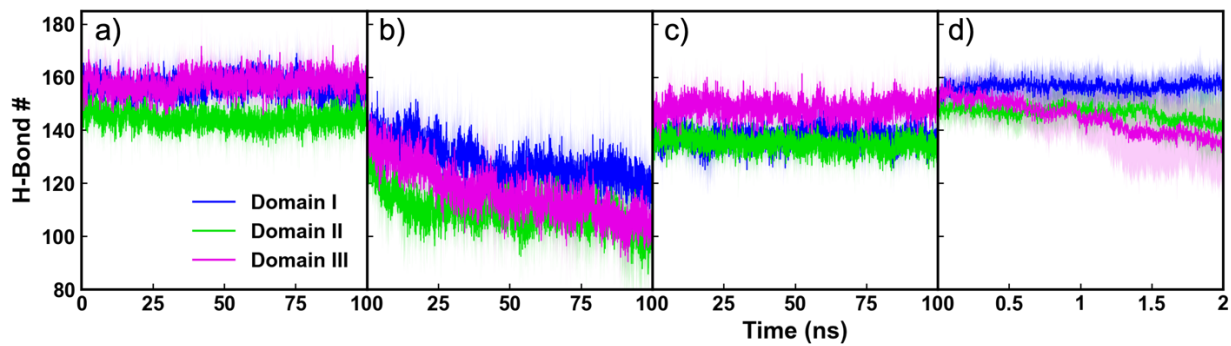

**Figure S5: Number of intra-domain hydrogen bond at (a) 310 K pH7.4 no shear, (b) 500 K pH7.4 no shear, (b) 310 K pH3.5 no shear, (c) 310 K pH7.4 shear rate 5×10⁻⁵ fs⁻¹. Solid lines indicate the mean values averaged over three independent trials, while the shaded regions show the standard deviation.**

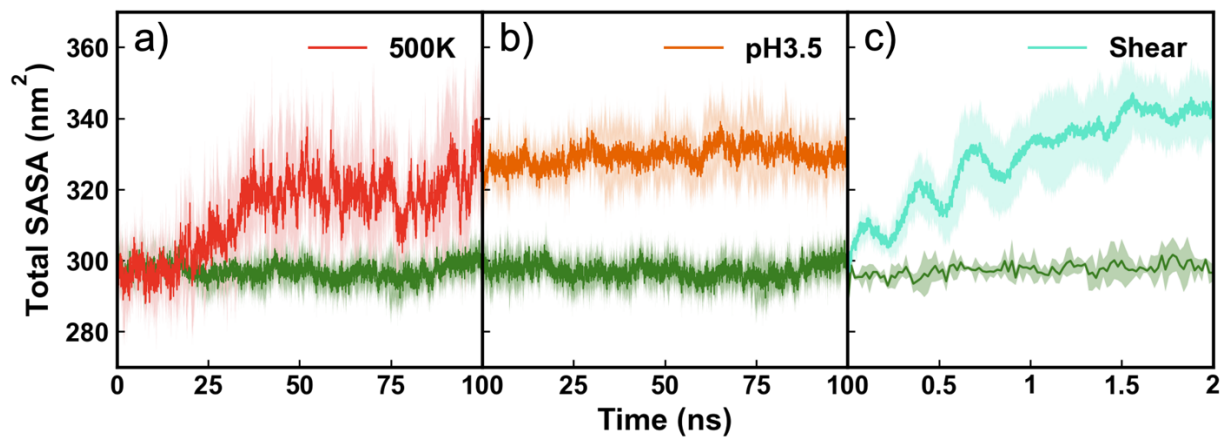

**Figure S6:** Total solvent accessible surface area (SASA) of BSA under (a) high temperature, (b) acidic pH, (c) shear stress. The solid lines indicate the mean values averaged over three independent trials, while the shaded regions show one standard deviation. We note that the maximum time for shear plots is 2 ns, whereas that for the high temperature and pH plots are 100 ns. (Deep green line as baseline taken from simulation at 310 K, pH = 7.4, and no shear applied).

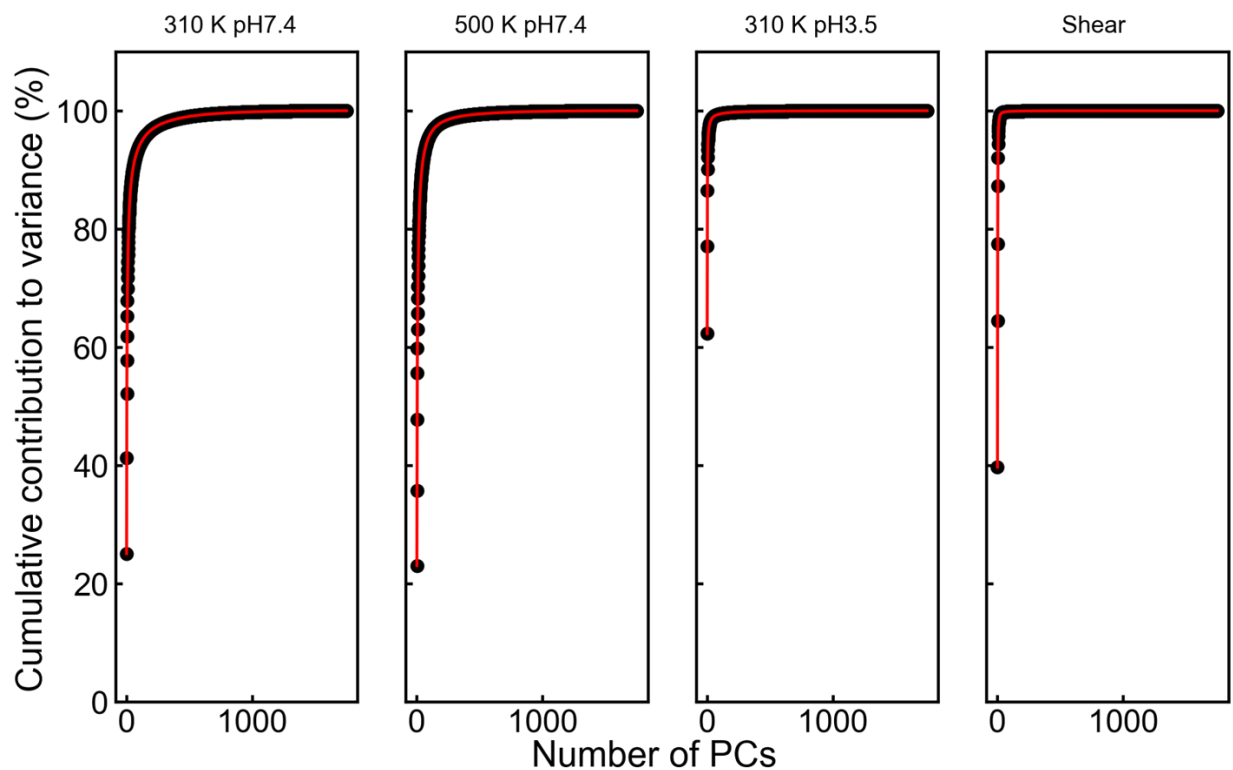

**Figure S7: Cumulative contribution (%, y-axis) of all the principal components (PCs, x-axis) to the variance of the overall BSA motions calculated upon Principal Component Analysis (PCA) under different conditions.**